\documentclass[aip,apl,reprint]{revtex4-1}
\usepackage{amsmath}
\usepackage{amsfonts}
\usepackage{amssymb}
\usepackage{graphics}
\usepackage{float}
\usepackage{graphicx}
\usepackage{epstopdf}
\usepackage[compact]{titlesec}
\usepackage{natbib}
\usepackage{threeparttable}
\usepackage[titletoc,title]{appendix}
\usepackage{lipsum}
\usepackage{varwidth}
\usepackage{tabularx}
\newcommand{\etal}{\textit{et al}. }

\begin{document}
\title{Imparity in valley population for single layer transition metal dichalcogenides under elliptical polarization}
\author{Parijat Sengupta}
\affiliation{Dept. of Electrical Engineering, University of Illinois at Chicago, Chicago, IL 60607.}
\author{Yaohua Tan}
\affiliation{Department of Electrical Engineering, University of Virginia, Charlottesville,
VA 22903}
\author{Dimitris Pavlidis}
\affiliation{Dept. of Electrical Engineering, Boston University, Boston, MA 02215.}
\author{Junxia Shi}
\affiliation{Dept. of Electrical Engineering, University of Illinois at Chicago, Chicago, IL 60607.}

\begin{abstract}
The illumination of a single-layer transition metal dichalcogenide with an elliptically-polarized light beam is shown to give rise to a differential rate of inter-band carrier excitation between the valence and conduction states around the valley edges, $ K $ and $ K^{'} $. This rate with a linear dependence on the beam ellipticity and inverse of the optical gap manifests as an asymmetric Fermi distribution between the valleys or a non-equilibrium population which under an external field and a Berry curvature induced anomalous velocity results in an externally tunable finite valley Hall current. Surface imperfections that influence the excitation rates are included through the self-consistent Born approximation. Additionally, we show that circular dichroism in the vicinity of the valley edges also exhibits an ellipticity dependence.
\end{abstract}
\maketitle

\begin{figure}[t!]
\centering
\includegraphics[width=0.9\linewidth]{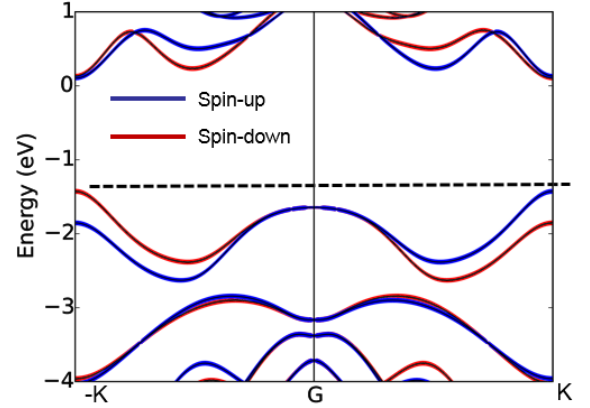}
\vspace*{-2mm}
\caption{The dispersion of mono-layer WS$_{2}$ as obtained from an ab-initio (VASP) calculation. The choice of WS$_{2}$ is dictated by the fact that it has a significant spin-orbit coupling allowing a clear distinction between the spin-split bands and the time reversal symmetry mandated order flipping at the valley edges. See supplementary material for details.}
\label{mldp}
\vspace{-1.8em}
\end{figure}

The valley Hall effect (VHE) or valleytronics~\cite{mak2014valley,mak2016photonics} as it is commonly known utilizes the valley degree of freedom to give rise to a non-dissipative current deemed useful for information processing and future memory designs. The  VHE, in principle, can manifest in materials that exhibit a valley structure but are most easily demonstrable in hexagonal lattices with a broken spatial inversion symmetry~\cite{wang2013valley}. The single layer transition metal dichalcogenides \textit{MX$_{2}$} (M = Mo, W; X = S, Se) which belong to this class of materials are characterized by two degenerate but inequivalent valleys $ K $ and $ K^{'} $. The two valleys which are time-reversed copies of each other (see Fig.~\ref{mldp}) carry an opposite and an out-of-plane directed orbital magnetization,~\cite{zhu2014study} an outcome of the broken symmetry induced Berry curvature~\cite{sundaram1999wave,sengupta2015influence}. The presence of such an orbital magnetic moment ensures the coupling of the valley pseudospin (the two valleys because of their opposite attributes are identified as pseudo-spins) to a magnetic field enabling an exclusive absorption of left- or right-circularly polarized light~\cite{cao2012valley,mak2012control} and in turn inducing optical transitions between eigen states whose azimuthal quantum numbers differ by unity. This strong optically governed valley polarization excites carriers which also acquire a Berry curvature imparted transverse anomalous velocity~\cite{xiao2010berry} thus furnishing a Hall-like current in presence of an external electric field. 

The valley Hall effect and the concomitant valley Hall current is remarkably robust attributed to a strong out-of-plane spin-splitting of valley states into a Kramers doublet (the spin-up state $ S_{z} = \hbar/2 $ at the $ K \left(\tau = 1\right) $ valley and the spin down $ S_{z} = -\hbar/2 $ at $ K^{'} \left(\tau = -1\right)$ are separated from their opposite spin counterparts by the spin-orbit coupling energy) that leads to an intrinsic spin-valley locking impairing valley relaxation via a simultaneous spin-flip and momentum conservation exchange between the $ K $ and $ K^{'} $ carrier ensembles. The valley Hall current is sensed as a voltage between two contacts via the inverse valley Hall effect~\cite{xiao2012coupled} and serves as the cornerstone of all theoretical predictions and experiments conducted heretofore. In general it is expressed as $ \overrightarrow{j} = ne\overrightarrow{v}_{an} $, where $ n $ is the carrier density and $ \overrightarrow{v}_{an} $ is the Berry curvature controlled anomalous velocity; likewise an equivalent expression (see also Eq.~\ref{ivhe}) in terms of the Hall conductivity $\left(\sigma_{xy} \right)$ and an electric field~\cite{li2012longitudinal} could be $ \overrightarrow{j} = \sigma_{xy}\overrightarrow{E} $. However, the intrinsic valley Hall current contribution from the edges, $ K $ and $ K^{'} $, are equal and oppositely directed yielding an overall vanishing current. The origin of this vanishing current is easy to see by noting (Eq.~\ref{ivhe}) that the Hall conductivity integral is the multiplicative product of the valley-specific Berry curvature $\left(\Omega\right)$ and Fermi distribution; for an identical valley Fermi distribution and the time reversal symmetry mandated relation $ \Omega\left(K\right) = -\Omega\left(K^{'}\right) $, the two edge conductivity tensor components cancel out. 
\begin{equation}
\sigma_{xy}^{\tau} = \dfrac{2e^{2}}{\hbar}\int\dfrac{\mathbf{d^{2}k}}{4\pi^{2}}f^{\tau}\left(k\right)\Omega^{\tau}\left(k\right). 
\label{ivhe}
\end{equation}
In Eq.~\ref{ivhe}, $ \Omega^{\tau}\left(k\right) = i\triangledown_{k} \times \langle u\left(k\right)\vert \triangledown_{k} \vert u\left(k\right)\rangle $ is the Berry curvature, $ u\left(k\right)$ are the Bloch functions while $ f^{\tau}\left(k\right) $ is the Fermi distribution in the vicinity of the valley edge.    

For a finite valley Hall current, the expression $ \sum_{\tau}\sigma_{xy}^{\tau} $ must be non-zero; the quantum of current and its modulation therefore hinging on creation of a non-equilibrium valley population, a condition simply represented as $ f^{K}\left(k\right) \neq f^{K^{'}}\left(k\right) $. The establishment of such an imbalance (in Fermi distribution) through a differential rate of excitation of carriers from the two valleys through elliptically polarized light is the central theme of this letter. Note that the valley edges show an exclusive absorption of left- or right-circularly polarized light. We find that the differential rate linearly depends on the ellipticity of the incident beam and inversely to the optical gap offering an external control over the measure of the valley Hall current. Carriers in single layer TMDCs may also experience an extrinsic side-jump endowing them with an extrinsic valley Hall conductivity directly connected to the difference in valley carrier distribution~\cite{xiao2007valley} as $ \sigma_{xy}^{vh^{'}} \approx \hbar^{2}\pi\Delta n/2m^{*}E_{g} $ (in units of e$^{2}$/h). The density difference between the photo-excited carriers in the two valleys is $ \Delta n = n_{K} - n_{K^{'}} $, the carrier effective mass is $ m^{*} $, and $ E_{g} $ denotes the valley band gap.

To model the carrier distribution inequality through inter-band transitions, we begin by noting the minimal Hamiltonian (Eq.~\ref{mH}) that describes the valleys. In its simplest form, the Hamiltonian is~\cite{xiao2012coupled}
\begin{equation}
H_{\tau} = at\left(\tau\,k_{x}\sigma_{x} +  k_{y}\sigma_{y}\right)\otimes\mathbb{I} + \dfrac{\Delta}{2}\sigma_{z}\otimes\mathbb{I} - \dfrac{\lambda\tau}{2}\left(\sigma_{z} - 1\right)\otimes\,s_{z}.
\label{mH}
\end{equation}
This $ 4 \times 4 $ Hamiltonian describes two non-interacting $ 2\times 2 $ blocks where the upper (lower) block furnishes the dispersion of the spin-up (down) conduction and valence bands. The lattice constant is $ a $ and $ t $ denotes the hopping parameter. The energy gap between the conduction and valence bands in absence of intrinsic
spin-orbit coupling is $ \Delta $. The Pauli matrices $ \hat{\sigma}_{i} $ where $ i = \left\lbrace x, y, z\right\rbrace $ act on the lattice sub-space while $ \hat{s}_{z} $ is linked to the spin of the electrons at valley edges $ K $ and $ K^{'} $. The valence (VB) and conduction band (CB) wave functions obtained through a direct diagonalization of the upper $ 2 \times 2 $ block in Eq.~\ref{mH} are $ \Psi_{VB} = 1/\sqrt{2}\begin{pmatrix}
\eta_{-}e^{-i\theta} & -\,\eta_{+} \end{pmatrix}^{T} $ and $ \Psi_{CB} = 1/\sqrt{2}\begin{pmatrix}
\eta_{+}e^{-i\theta} & \,\eta_{-} \end{pmatrix}^{T} $, where $ \theta = arg\left(atke^{-i\tau\delta}\right) $, $ \delta = \tan^{-1}k_{y}/k_{x} $ and the pair of coefficients $ \eta_{\pm} = \sqrt{1 \pm \alpha} $. The parameter $ \alpha $ is defined as
\begin{equation}
\alpha = \dfrac{\Delta - s\tau\lambda}{\sqrt{\left(\Delta - s\tau\lambda\right)^{2} + \left(2atk\right)^{2}}}.
\label{vbcoeff}
\end{equation}
Here $ s = 1 $ points to the upper $ 2\times 2 $ spin-up block. The wave function expressions for the identical sized lower block for the spin-down bands can be simply written by setting $ s = -1 $ in Eq.~\ref{vbcoeff}. The  CB (+) and VB (-) eigen energies are $ \varepsilon_{s,\pm} = \dfrac{1}{2}\left[s\lambda\tau \pm \sqrt{\left(\Delta - s\lambda\tau\right)^{2} + \left(2atk\right)^{2}}\right] $. Dispersion plots within the \textit{k.p} representation for semiconducting single layer TMDCs are included in supplementary material. Notice that the $ 4 \times 4 $ structure of the Hamiltonian furnishes two inter-band transition gaps; each gap is measured from top of the two spin-split VB to the spin degenerate CB minimum. The two inter-band transition energies are $ \sqrt{\left(\Delta \mp \lambda\tau\right)^{2} + \left(2atk\right)^{2}} $. The upper (lower) sign corresponds to gap between the spin-up (down) CB and VB bands. For all calculations, we tacitly assume that a transition happens from top of the highest valence band to bottom of conduction band; to that end, $ s $ is always set to +1 (-1) for $ K \left(K^{'}\right) $.

A direct computation of the inter-band transition matrix element (or the transition probability of carrier excitation from top of VB to bottom of CB) with the aforementioned wave functions offers a pathway to the quantitative determination of carrier imbalance in the two valleys. Note that two valleys absorb light differently and the carrier excitation rates differ leading to the predicted dissimilarity in their respective Fermi distribution. For our purpose, we consider an elliptically polarized light that shines on the surface of a single layer TMDC and serves as the external agent for inter-band transitions. The electric field in the most general form for an arbitrarily polarized light beam is $ \vec{\mathbf{E}} = E_{x}\hat{x}\cos\left(kz - \omega t\right) +  E_{y}\hat{y}\cos\left(kz - \omega t + \phi\right) $. For calculations that follow, we set $ \phi = \pi/2 $ and $ \vert E_{x} \vert = \vert E_{y} \vert/\mu = \vert E_{0} \vert $, where $ \mu \neq 0 $ denotes the ellipticity ratio of the beam. The influence of the electric field through the corresponding magnetic vector potential $ \overrightarrow{A} $, which are linked by the relation $ \overrightarrow{E}(t) = -\left(1/c\right)\partial\overrightarrow{A}/\partial t $, is included in the Hamiltonian using the standard Peierls substitution, $ \overrightarrow{p} =  \overrightarrow{p} - \left(e/c\right)\overrightarrow{A} $. The speed of light is indicated by $ c $ in appropriate units. Choosing the upper block of the Hamiltonian (Eq.~\ref{mH}) and substituting for $ \overrightarrow{p} $ gives the interaction Hamiltonian (see supplementary material) as $ \mathcal{\hat{H}}_{int} = \left(ie/2\hbar\omega\right)at\left(\tau\sigma_{x}E_{x} +  \sigma_{y}E_{y}\right) $. The factor of 0.5 comes from retaining only the $ (-i\omega t)$ term while expressing the vector potential in the complex plane. For our case, $ E_{x}\hat{x} = \vert E_{0} \vert/\sqrt{1 + \mu^{2}}\hat{x} $ and $ E_{y}\hat{y} = i\mu\beta\vert E_{0} \vert/\sqrt{1 + \mu^{2}}\hat{y} $, where $ \beta = 1 \left(-1\right) $ identifies the right (left) elliptically polarized light. With this in mind, the inter-band matrix element therefore can be written as $ \mathcal{M} = \langle \Psi_{VB}\vert \mathcal{\hat{H}}_{int} \vert \Psi_{CB}\rangle $. Explicitly evaluating the matrix element using $ \mathcal{\hat{H}}_{int} $ at $ K $ and $ K^{'} $, we have
\begin{flalign}
\vert \mathcal{M}_{\tau} \vert^{2} = \Upsilon\biggl[\left(\tau\alpha + \mu\beta\right)^{2}\cos^{2}\theta + \left(\tau + \alpha\mu\beta\right)^{2}\sin^{2}\theta \biggr].
\label{matelm}
\end{flalign} 
The coefficient $ \Upsilon = \left(1 + \mu^{2}\right)^{-1}\left(e a t E_{0}/2\hbar\omega\right)^{2} $ is a function of the electric field (and power) of the incident light beam and material parameters for the selected single layer TMDC. 

The difference in photo-excitation rates or the carrier lifetime at $ K $ and $ K^{'} $ can now be computed by inserting the square of the matrix element (from Eq.~\ref{matelm}) in the Fermi golden rule rate expression. Summing over available momentum states $ \left(\sum k = \left(1/4\pi^{2}\right)\int kdk\int d\theta\right) $ in a cut-off range close to the valley edge , we obtain the following rate equation:
\begin{flalign}
\dfrac{\partial n}{\partial t} &= \dfrac{\Upsilon}{2\hbar}\int kdk \int_{0}^{2\pi}d\theta\left[\vert \mathcal{M}_{K} \vert^{2} -  \vert \mathcal{M}_{K^{'}} \vert^{2}\right]\delta\left(\varepsilon - \hbar\omega\right), \notag \\
&= \dfrac{\Upsilon}{8a^{2}t^{2}\hbar}\left[\left(\vert \mathcal{M}_{K}^{'}\left(\mathcal{E}\right)\right)\vert^{2} - \left(\vert \mathcal{M}_{K^{'}}^{'}\left(\mathcal{E}\right)\right)\vert^{2}\right].
\label{fgrdiff}  
\end{flalign}
In Eq.~\ref{fgrdiff}, $ \vert \mathcal{M}_{\tau}^{'}\left(\varepsilon\right)   \vert = \left(2\varepsilon - s\tau\lambda\right)\vert \mathcal{M}_{\tau}\left(\varepsilon\right) \vert $; also note that the angular part is integrated out to replace the sinusoidal terms by $ \pi $. To change the variable of integration from momentum to energy space, we have rewritten $ \alpha $ as  $ \left(\Delta - s\tau\lambda\right)/\left(2\varepsilon_{+} - s\tau\lambda\right) $ using Eq.~\ref{vbcoeff}. The Fermi level is set to top of the highest filled spin-up (down) valence band at $ K \left(K^{'}\right)$ and a photo-transition happens when a carrier is excited to the empty conduction band. The difference in energy between the VB and CB states is exactly equal to $ \hbar\omega $. 

The differential rate of excitation strictly at the valley edges is easy to determine; setting $ \vert k \vert = 0 $ in Eq.~\ref{fgrdiff} yields a simple expression : $ \partial n/\partial t = \left(e^{2}\vert E_{0} \vert^{2}/4\hbar\Delta^{'}\right)\mu\beta  $. In obtaining this rate equation, we have used the relation $ \Delta^{'} = \Delta - \lambda = \hbar\omega $, the inter-band transition (optical) gap between the top of highest valence band and bottom of conduction band. At this point, it is instructive to ascertain if the derived rate expression is in accord with the experimental observation that reports the exclusive absorption of circularly polarized light of a distinct chirality at the valley edges. To perform such a check using Eq.~\ref{matelm}, we let $ \mu = 1 $ for a circularly polarized light beam, $ \beta = 1 $ for right-handedness and $ \alpha $ goes to unity since $ \vert k \vert = 0 $ at the valley edge. Substituting these values in Eq.~\ref{matelm}, it is straightforward to see that $ \vert M \vert = 0 $ for the $ K^{'} $ valley $\left(\tau = -1\right)$ as expected. Similarly, setting $ \beta = -1 $ gives a vanishing matrix element for the $ K \left(\tau = 1\right) $ valley edge. 

The derived rate at the valley edge shows an explicit dependence on the band gap suggesting that a modulation of it could serve as an effective control mechanism. Experimental data about the photo luminescence spectra of single layer TMDCs reveal peaks which correspond to a band gap marked by a distinct influence of temperature similar to that observed in conventional semiconductors. While an exact formulation of this microscopic relationship is manifestly a many-body phenomenon, a semi-empirical relation between band gap and temperature, first proposed~\cite{o1991temperature} by O'Donnel \etal  confirms that indeed for a rise in temperature, an enhancement in carrier excitation rate is observed. The supplementary material describes this in greater detail.
 
\begin{figure}[!t]
\includegraphics[scale=0.85]{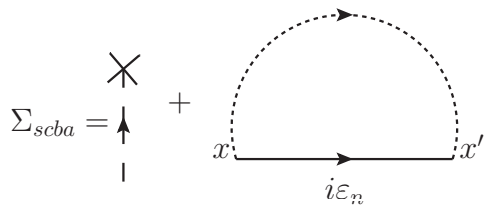} 
\vspace*{-2mm}
\caption{The self energy $ \left(\Sigma_{scba}\right) $ in the Born approximation averaged over impurity distributions. The Matsubara frequency is unchanged since collisions are assumed to be elastic. The dashed line is the average of the two impurity locations marked as $ x $ and $ x^{'} $ while the $ \times $ represents a scattering event.}
\label{feyn1}
\vspace{-2.2em}
\end{figure}

In addition to temperature adjusted energy gap, we also wish to draw attention to the presence of surface imperfections and disorder that are inevitably present and reveals as a Lorentzian broadening. The density of states, $ g\left(\varepsilon\right) $, expressed through the inclusion of the $ \delta\left(\cdot\right) $ function in Eq.~\ref{fgrdiff} is now represented as $ g\left(\varepsilon\right) = \left(1/\pi\right)\left[\Gamma/\left( \varepsilon^{2} + \Gamma^{2}\right)\right] $. The broadening $\left(\Gamma\right)$ can be straightforwardly computed within the self-consistent Born approximation~\cite{koshino2006transport} (SCBA). Assuming a distribution of random potential scattering nodes $ \left(\hat{V}\left(\hat{r}\right)\right) = \sum_{i}\xi_{i}\delta\left(\vec{r}-\vec{R}_{i}\right) $ that impede electronic motion, the SCBA as a perturbative calculation lets us model this disorder as a self-energy $\left(\Sigma\right)$ through the Dyson equation $ \hat{G}^{-1} = \hat{G}^{-1}_{0} - \hat{\Sigma} $. The broadening is extracted from the imaginary part of the self-energy which consists of two terms, the single and double scattering from an impurity (see Fig.~\ref{feyn1}). The amplitude of the potential at each such scattering node is denoted by $ \xi $ and supposed to be of identical strength. Note that $ \hat{G}\left(i\omega_{n}\right)$ and  $ \hat{G}_{0}\left(i\omega_{n}\right) $ represent the finite-temperature disordered and bare Matsubara Green's function~\cite{bruus2004many}. The fermionic Matsubara frequencies are $ \omega_{n} = \left(2n +1\right)\pi kT $. To a lowest order approximation, the Dyson equation gives the disorder-induced self-energy as $ \Sigma = \xi/4\pi^{2}\int\,d^{2}\vec{k}\,G_{0}\left(i\omega_{n},\vec{k},\varepsilon\right) $. The supplementary material includes a numerical calculation of the broadening for a preset impurity potential and concentration~\cite{sengupta2016photo}. 

\begin{figure}[!t]
\centering
\includegraphics[scale=0.6]{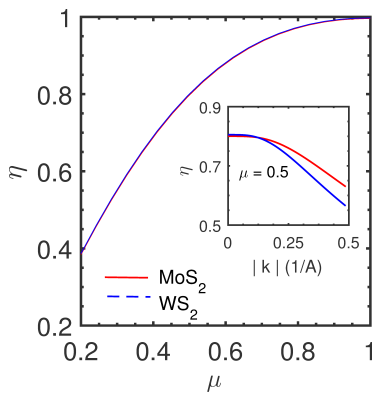} 
\vspace*{-3mm}
\caption{The circular dichroism in the vicinity of the valley edges is plotted for several ellipticity ratios. For this plot, we have set $ \vert k \vert = 0.1 1/\AA $. Note that as ellipticity approaches unity or full circular polarization, $ \eta $ is $ \approx 1 $. The inset depicts (for $ \mu = 0.5 $) $ \eta $ for a set of $ \ k $ values around the valley as it tails off akin to the diminishing Berry curvature of the Bloch bands with increasing distance from the edge.}
\label{cdmu}
\vspace{-2em}
\end{figure}

We briefly digress here to note that a particularly useful expression about circular dichroism $\left( \eta \right)$ which is the preferential absorption of left- or right-circularly polarized light and has applications in the design of meta-materials -primarily in areas of polarization sensitive imaging devices and display technologies- can be easily obtained from the above results. Note that circular dichroism which is uniquely determined for single layer TMDCs is a direct consequence of valley orbital magnetic moment of the Bloch electrons located here. Using notation from previously published result~\cite{ezawa2012spin,sengupta2016tunable}, we define it as a fraction of the difference in polarization-dependent to the total absorption. It takes the form:
\begin{equation}
\eta\left(k\right) = \dfrac{\vert
\mathcal{P}_{+}\left(k\right)\vert^{2}-\vert
\mathcal{P}_{-}\left(k\right)\vert^{2}}{\vert
\mathcal{P}_{+}\left(k\right)\vert^{2}+\vert
\mathcal{P}_{-}\left(k\right)\vert^{2}}, 
\label{cd1}
\end{equation}
where the quantities $\mathcal{P}_{\pm}\left(k\right)$ are described in terms of the inter-band matrix elements as $\mathcal{P}_{\pm}\left(k\right)\,=\,\mathcal{M}_{cv}^{x}\,\pm\,i\mathcal{M}_{cv}^{y}$. The inter-band matrix elements are expressed in the usual way, for instance, $\mathcal{M}_{cv}^{x}$ is given by $ \mathcal{M}_{cv}^{x} = \langle\Psi_{CB}\vert \hat{v}_{x}\vert \Psi_{VB}\rangle $. Utilizing the result from Eq.~\ref{matelm} and substituting in Eq.~\ref{cd1}, the circular dichroism as a function of the ellipticity and location in momentum space is 
\begin{equation}
\eta = \int_{0}^{2\pi}d\delta\dfrac{\beta\Delta^{'}\mu\tau{\sqrt{\Delta^{'2} + \left(2atk\right)^{2}}}}{g\left(k\right)\left(1 + \mu^{2}\right) - 2\left(atk\right)^{2}\left(1 - \mu^{2}\right)\cos2\delta}.
\label{cd2}
\end{equation}
In Eq.~\ref{cd2}, we have defined $ g\left(k \right) = \Delta^{'2} + 2\left(atk\right)^{2} $, $ \Delta^{'} = \left(\Delta - s\tau\lambda\right)/\pi $, and $ \delta = \tan^{-1}k_{y}/k_{x} $. The circular dichroism at the valley edges must be $ \pm 1 $; a quick check by setting $ \vert k \vert = 0 $ and $ \mu = 1 $ (for circularly polarized light) reduces the integral in Eq.~\ref{cd2} to $ \eta = \tau\beta = \pm 1 $, confirming the validity of the expression. We plot this dependence of circular dichroism on the ellipticity (Fig.~\ref{cdmu}) for two prototypical single layer TMDCs, Mo$S_{2}$ and W$S_{2}$ with similar band parameters. The valence band spin-splitting in W$S_{2}$ is however, approximately three-fold higher, the heavy metal Tungsten being the major contributor.
 
\begin{figure}[!t]
\centering
\includegraphics[scale=0.6]{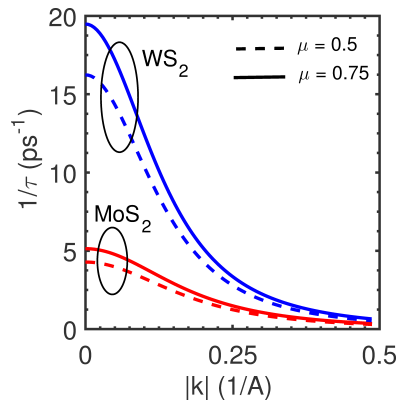} 
\vspace*{-2mm}
\caption{The differential rate of excitation $\left(1/\tau\right) $ for Mo$S_{2}$ and W$S_{2}$ is shown as a function of distance from the valley edge. The rates were computed at $ T = 300 \, K $ and include an impurity-induced broadening of the DOS.}
\label{diffr}
\vspace{-1.9em}
\end{figure}

For a numerical estimate of carrier excitation rate, we again select the TMDCs Mo$S_{2}$ and W$S_{2}$. The impurity created broadening of the DOS in each case was computed to be $ 8.0 \,meV $ and $ 4.6 \,meV $, respectively (see supplementary material). Inserting the appropriate band parameters and the Lorentzian broadening for the DOS, the differential excitation rates for equi-energetic surfaces around the valley edges is shown in Fig.~\ref{diffr}. Note how the differential rate progressively drops as we include points in momentum space far away from the valley edges; at such points the topologically governed out-of-plane orbital magnetic moment that necessitates the exclusive polarization-sensitive absorption is lost. The plots for a pair of ellipticity ratios $\left(\mu\right) $ also underscore the fact that for a higher ellipticity which imparts an enhanced `'circular'' polarization character, the differential rate is boosted in agreement with the phenomenon of preferential absorption. It is worthwhile to note that applications which rely on a definite measure of valley current may therefore be tuned by a simple alteration of $ \mu $, commonly achieved by letting a circularly polarized beam pass through a linear polarizer. 

The production of non-equilibrium carrier density through elliptical pumping can be initiated in any hexagonal lattice with a valley structure; however, fabrication issues notwithstanding and the absence of high-quality TMDC crystals, the large band gap at the valleys in the exfoliated single layer when compared to dual valley graphene makes them a more promising candidate for easier valley polarization with commercially available light sources. Furthermore, similar to electrons, trions that carry charge and are known to exist in single-layer TMDCs~\cite{lui2014trion} as long-lived excitations can also acquire a Berry curvature and observe the valley optical selection rule making them conformable to the elliptical pumping controlled rate analysis presented in this letter.

%\bibliographystyle{apsrev}
%\bibliography{reference} 

\end{document}